\documentclass[final,3p,times]{elsarticle}

\usepackage{epsfig}
\usepackage{makeidx}
\usepackage{graphicx}
\usepackage{epstopdf}

\usepackage{amssymb}

\biboptions{}

\def\pb{\, .}
\def\vb{\, ,}

\begin{document}

\begin{frontmatter}

\title{From microscopic taxation and redistribution models \\
to macroscopic income distributions}


\author{Maria Letizia Bertotti\corref{cor1}}
\ead{MariaLetizia.Bertotti@unibz.it}
 
\author{Giovanni Modanese}
\ead{Giovanni.Modanese@unibz.it}
 
\cortext[cor1]{Corresponding author}

 \address{Faculty of Science and Technology,
   Free University of Bozen-Bolzano \\
   Piazza Universit\`a 5, 39100 Bolzano, ITALY}



\begin{abstract}
We present here a general framework, 
expressed by a system of nonlinear differential equations,
suitable for the modelling of 
taxation and redistribution
in a closed 
society.
This framework  allows
to describe the evolution of the income distribution over the population
and to explain the emergence of collective features
based on the knowledge of the 
individual interactions. 
By making different choices of the framework parameters,
we construct different
models, 
whose long-time behavior is then investigated. 
Asymptotic stationary  distributions are found, which 
enjoy
similar properties as those observed in empirical distributions.
In particular, they exhibit
power law tails of Pareto type
and their Lorenz curves and Gini indices
are consistent with some real world ones.
\end{abstract}

\begin{keyword}
Econophysics; Taxation and redistribution model; Income distribution; Power law; Pareto tail
\end{keyword}

\end{frontmatter}

\section{Introduction}

The interest of physicists and mathematicians towards  
complex systems arising in social and economical sciences
has been
constantly growing in the last years, as attested
by the number of published papers. 
Among the subjects these papers deal with, 
one finds 
opinion formation dynamics (see for example \cite{BNa Kra Vaz Red, Ber Del 1, Ber Del 2, Gra Kos, Wei Def Amb}),
relaxation processes to
steady wealth and income distributions \cite{Cha Cha, ChaFig Mou Rib, Chatt Cha, Cle Gal, Dur Mat Tos 1, Dur Mat Tos 2, Pat},
mechanisms of financial markets and other out-of-equilibrium economic
and financial phenomena \cite{Art, Man Sta, Sin Cha, Yak}.
These topics share the common fact of 
referring to systems (populations) composed by a large number of interacting elements (individuals).
And this is 
why methods and tools
from statistical mechanics and kinetic theory have been and are being adapted and employed
to investigate them.

In 
a recent paper,
\cite{Ber}, one of the present authors introduced a general framework,
suitable for the construction of models of taxation and redistribution 
in a closed society.
This framework originates from a discrete active particle kinetic approach \cite{Bel} and
is expressed by
a system of nonlinear ordinary differential equations. 
The equations are
as many as the classes, 
each one characterized by its \textquotedblleft average income\textquotedblright, in which the population can be divided.
Each equation gives the variation in time of the fraction of individuals belonging to a certain class.
The framework provides a description of the evolution of the
wealth\footnote{\ The words income and wealth are used in this paper to design the same concept.} distribution over the population
and aims at explaining emerging collective features,
based on the knowledge of the individual interactions.
As a case study, also a specific model was formulated in \cite{Ber}. To this end, the general mathematical framework
was exploited and a particular choice 
for the values of some of its parameters was made.
The well-posedness of the model as well as the existence of two conserved quantities, corresponding to 
the total population and the global wealth,
was then 
established.
Several simulations were carried out for the case in which the number $n$ of income classes (and of differential equations)
is equal to $5$.
Specific attention was devoted in \cite{Ber} to the differences,
detectable
from the shape of the long-time income 
distributions, 
among systems with different taxation rates.
The result was that increasing 
the difference between the maximum and the minimum tax rate
leads, at the asymptotic equilibrium, to the growth of the middle classes, 
to the detriment of the poorest and the richest classes. 
This is a reasonable feature which
encourages,
in spite of the naiveness 
and roughness
of the model,
a thorough investigation. 
A natural observation is that
the study of the model in the particular case
with 5 income classes does not allow for example to recover 
the Pareto-law which is observed in real world economies. 
As a matter of fact,
$n=5$ is too small a number for 
a tail of the steady income distribution to show up.
In view of that, we started
performing
a great deal of computational simulations relative to the model with greater values of $n$.
We tried and considered several choices of the model parameters,
expressing e.g.
different characterizations of the incomes or of the tax rates.
With the aim of treating reasonable cases,
as far as possible comparable with real world ones,
we focused our attention on 
\textquotedblleft realistic\textquotedblright 
initial population distributions,
where the majority of individuals belongs to lower income classes,
while higher classes are less densely populated (see Section $3$ for details).
The numerical solutions systematically show that
for any fixed value of the total wealth
a unique asymptotic stationary distribution exists,
independently of the random choice of the initial population distribution 
(subjected only to the just mentioned 
\textquotedblleft realistic\textquotedblright requirement).
The asymptotic stationary distribution exhibits the following patterns:
the density of the low income classes is smaller than for the low-medium classes,
the maximal density is achieved by the low-medium classes
and the density 
progressively
decreases for the higher income classes.
In fact, the asymptotic distributions exhibit salient features of empirical distributions 
(see e.g. \cite[page $14$]{Bancad'Italia08}, \cite[page $19$]{Bancad'Italia10}, and
\cite[page $8$]{ISTAT}).
At a closer look, one also finds that the tails of the distributions
have indeed a power law behavior \cite{New}.
An analysis of the 
shape of the tails
and its relation with the parameters of the model
is the subject
of the present paper.

\medskip

The paper is organized as follows.
In Section $2$, we review for the convenience of the reader the framework and the model introduced in \cite{Ber}.
Section $3$ focuses on the existence of asymptotic stationary distributions and on their properties. In particular, Pareto tails \cite{Par} are found to
occur and their Pareto indices are calculated in some different cases.
The last section provides a short summary of the content of the paper and a brief mention of possible future developments.

\medskip

Before starting, it may be of interest comparing certain features of our models with those of others available in the econophysics or 
\textquotedblleft classical\textquotedblright mathematical economics literature. 
Some immediately evident differences concern the mathematical formulation of the income distribution problem. 
A first class of works (see \cite{Pat} for a review) develop a statistical analysis of the population by means of Montecarlo simulations 
of the interactions of a large number of individuals. In these models the interaction rules can be defined with great freedom, because they are applied 
in a straightforward way through the simulation algorithm. 
At the same time, the method is not based upon evolution equations. Therefore it lacks general mathematical theorems 
which should keep account of specific hypotheses on the interactions. 
To draw a comparison, within our framework we are free to fix several parameters (interaction frequency, taxation rates, etc.), 
but 
for the models to be conservative,
some terms in the transition probabilities expressing income class changes
must be proportional to the reciprocal of the income difference $|r_i-r_j|$. If one changes this dependence, 
supposing for instance that the transition probability is proportional to the reciprocal of $|r_i-r_j|^2$, then the 
conservation of the total wealth
ceases to be valid. 
The dependence of the model on the details of the interactions is 
also typical of the approaches based on the Boltzmann equation \cite{Dur Mat Tos 1, Dur Mat Tos 2}.
In the application of this equation to the kinetic theory of gases the interactions are determined to a large extent by physical conservation laws and symmetry principles. 
In the applications to econophysics there is some more freedom as for the definition of the interactions. 
Within this approach it is possible proving powerful general results concerning the moments of the distribution function $f(w,t)$ and the Pareto index
of the stationary asymptotic distribution. It has to be noticed however that
the structure of the Boltzmann equation 
requires advanced mathematical tools for his treatment:
the partial time derivative of the income distribution function $f(w,t)$ is given by an integral operator 
acting on $f(w,t)$ and also involves an average on some stochastic variables. 
The time evolution is usually computed through some approximate discretization method. Similar approximations are also employed 
in the classical economics theory \cite{Hee Mau}. There, the time evolution equations are not derived 
from hypotheses on the \textquotedblleft microscopic\textquotedblright interactions, but from variational principles 
and other general \textquotedblleft macroscopic\textquotedblright considerations.
Of course, the classical approach 
is characterized by a more realistic 
description of the dynamics of a complex
economy,
taking into consideration also
different kinds of assets, financial transactions, government intervention etc..
Our framework introduces 
from the beginning a discretization of the distribution function. We suppose that the individuals belong to income classes and pass from one class to another with certain probabilities. 
In our case, the equivalent of the classical distribution function could be written as a liner combination of Dirac delta-functions:
$$
f(w,t)=\sum_i x_i(t) \delta(w-r_i) \pb
$$
The quantities which evolve in time are discrete, like the individuals in a Montecarlo simulation, but in fact they 
express already averages, 
weighed with certain probabilities, much like the matrix elements of a wavefunction in quantum mechanics. The formalism is familiar to physicists, 
since it reminds of the Schroedinger equation of an atomic system expanded on a basis of states. 
This approach was previously devised to describe problems of opinion formation 
\cite{Ber Del 1, Ber Del 2}, for which the state of an individual can be reasonably represented by a discrete rather than real variable.
It provides some advantages, however, also for the treatment of income distributions. 
It allows, for instance, a very natural definition of different taxation rates for different income classes, and a division of the population in classes 
with incomes that increase non-linearly. In turn, this enables a better representation of the ''super-rich'' classes in a population.

\section{Taxation and redistribution in a closed trading market society}

In this section we shortly describe the general mathematical framework introduced in \cite{Ber}
for the modelling of the dynamical process of taxation and redistribution in a closed trading market society.
We then construct a particular family of models,
by attributing specific values to the parameters in the framework.

\subsection{A general framework}

Before writing down the system of nonlinear ordinary differential equations
which constitute the framework,
we 
briefly describe the context and introduce some notation.

Consider a population of individuals belonging to a finite number $n$ of classes,
each one characterized by its \textquotedblleft average income\textquotedblright. Let $r_1,r_2 \ldots r_n$
denote the average incomes
of the $n$ classes, ordered so that $r_1\le r_2 \le \ldots \le\ r_n$,
and let $x_i(t)$, where
$x_i : {\bf R} \to [0,+\infty)$ for $i \in \Gamma_n = \{ 1,2, ..., n \}$, denote the fraction at time $t$
of individuals belonging to the $i$-th class.
In the following, the indices $i, j, h, k$, etc. will always belong to $\Gamma_n$.

What produces the dynamics is a whole of pairwise interactions of economic nature, subjected to taxation.
Call $S$ the fixed amount of money that people may exchange during their interactions.

Any time an individual of the $h$-th class has to pay a quantity $S$ to an individual of the $k$-th class, 
this one in turn has to pay some tax corresponding to a percentage of what he is receiving.
This tax is quantified as $S \, \tau$, where the tax rate $\tau = \tau_k \le 1$ depends 
in general on the class of the earning individual. Since 
the amount of money corresponding to the tax $S \, \tau$ should go to the government, which
then is supposed to use the money collected through taxation to
provide welfare services  for the population, we 
interpret the welfare provision as an income redistribution.
Ignoring the passages 
to and from the government,
we 
adopt the following equivalent mechanism as the mover of the dynamics:
in correspondence to any interaction between an $h$-individual and a $k$-individual, where the one who has to pay $S$ to the 
other one is
the $h$-individual,
as a matter of fact
the $h$-individual pays to the $k$-individual a quantity 
$S \, (1 - \tau)$ and he pays as well a quantity $S \, \tau$, which is divided among
all $j$-individuals for $j \ne n$.\footnote{\ The reason why individuals of the $n$-th class constitute an exception is a technical one:
if an individual of the $n$--th class would receive some money, 
the possibility would arise 
for him to advance to a higher class, which is impossible.}
Accordingly, the effect of taxation and redistribution 
is equivalent to the effect of a quantity of interactions 
between the $h$-individual and each one of the $j$-individuals for $j \ne n$,
which are
\textquotedblleft induced\textquotedblright by the effective $h$-$k$ interaction.
To fix notations,
we
may distinguish between \textquotedblleft direct\textquotedblright interactions ($h$-$k$) and \textquotedblleft indirect\textquotedblright interactions ($h$-$j$
for $j \ne n$).

Now: any direct or indirect economical interaction
yields as a consequence a possible  slight increase
or slight decrease of
the income of individuals.

To translate in mathematical terms all that, we introduce

{$\bullet$} the (table of the) {\slshape interaction rates}
$$
\eta_{hk} \in [0,+\infty)  \vb
$$
expressing the number of effective encounters per unit time between
individuals of the $h$-th class and individuals of the $k$-th class;

{$\bullet$}  the (tables of the) {\slshape direct transition probability densities} 
$$
C_{hk}^i \in [0,+\infty)  \vb
$$
satisfying for any fixed $h$ and $k$
$$
\sum_{i=1}^n C_{hk}^i = 1  \vb
$$
which 
express the probability density that an individual of the $h$-th class 
will belong to the 
$i$-th class after a direct interaction with an
individual of the $k$-th class;

{$\bullet$}  the (tables of the) {\slshape indirect transition variation densities} 
$$
T_{[hk]}^i  \, :   {\bf R}^n \to {\bf R}  \vb
$$
where the $T_{[hk]}^i(x)$ with $x=(x_1, ..., x_n) \in {\bf R}^n$ are continuous functions, satisfying, for any fixed $h$, $k$ and
$x  \in {\bf R}^n$ 
$$
\sum_{i=1}^n T_{[hk]}^i(x) = 0 \pb
$$
These functions account for the indirect interactions and express the 
variation density in the $i$-th class
due to an interaction between an individual of the $h$-th class
with an individual of the $k$-th class.

\smallskip

If the interactions rates $\eta_{hk}$ are chosen for simplicity all equal to $1$,
the evolution of the class populations $x_i(t)$ 
is governed by the following differential equations, 
in which, of course, the contribution of both direct and indirect interactions is present:
\begin{equation}
{{d x_i} \over {d t}} =  
\sum_{h=1}^n \sum_{k=1}^n {\Big (} C_{hk}^i + T_{[hk]}^i(x) {\Big )}
x_h x_k     -    x_i  \sum_{k=1}^n x_k \pb
\label{evolution eq eta = 1}
\end{equation}

This is a 
system of nonlinear ordinary differential equations.
For instance, if  the values of the elements $T_{[hk]}^i$ are chosen as in the next subsection,
the r.h.s. of the equations $(\ref{evolution eq eta = 1})$
is a polynomial of third degree.

\subsection{A family of models: special choices for the transition probabilities}

In order to design within the general framework 
just discussed a specific model (or a specific family of models), we need to further characterize the expressions 
of the direct transition probability densities $C_{hk}^i$ 
and 
the indirect transition variation densities $T_{[hk]}^i(x)$. A conceivable choice is given next.

\smallskip

As done in the previous subsection, we take all the interaction rates $\eta_{hk}$ to be equal to one.
This corresponds to assuming that all the encounters between two individuals occur
with the same frequency, independently of the classes to which the two belong.

\smallskip

We represent the direct transition probability densities $C_{hk}^i$ as
$$
C_{hk}^i = a_{hk}^i + b_{hk}^i \vb
$$
where
the term $a_{hk}^i$ 
expresses the probability density that an $h$-individual
will belong to the $i$-th class
after an encounter with a $k$-individual,
when such an encounter does not produce any change
of class
and the term $b_{hk}^i$ expresses the 
density variation in the $i$-th class
of an $h$-individual interacting with a $k$-individual.

\smallskip

Accordingly, the only nonzero elements
$a_{hk}^i$ are
$$
a_{ij}^i = 1 \pb
$$

\smallskip

To define the elements $b_{hk}^i$,
we introduce the matrix $P$, whose elements $p_{h,k}$ 
express the probability that  in an encounter between an $h$-individual and a $k$-individual,
the one who pays is the $h$-individual. Admitting also the possibility that
to some extent the two individuals do not really interact, we have $0 \le p_{h,k} \le 1$ and, 
furthermore, $p_{h,k} + p_{k,h} \le 1$. 

There is some arbitrariness in the construction of the matrix $P$.
What is important is putting each element in the first row, as well as each element
but the very last one in the last column, equal to zero.
Indeed, since there is not a class lower than the first
nor one higher than the $n$-th, we cannot admit the possibility 
for $1$-individuals [respectively, for $n$-individuals] to move back to
a lower class [respectively, to advance, passing in a higher class]. Observing that also
in the presence of interactions between individuals of the same class, the average wealth of the class itself decreases
because of payment of taxes, we assume that individuals of class $1$ never pay 
(nor even $1$-individuals) while individuals of class $n$ can only receive money
from other $n$-individuals.

An encounter between an $h$-individual and a $k$-individual, with 
$h \ge 2$ and $k \le n-1$
and
the
$h$-individual paying, 
produces 
the elements 
\begin{eqnarray}
& b_{hk}^{h-1} = p_{h,k} \, S \, (1-\tau_k) \, \frac{1}{r_h - r_{h-1}} \vb 
\quad b_{hk}^h = - p_{h,k} \, S \, (1-\tau_k) \, \frac{1}{r_h - r_{h-1}} \vb \nonumber \\
& b_{kh}^{k+1} = p_{h,k} \, S \, (1-\tau_k) \, \frac{1}{r_{k+1} - r_{k}} \vb
\quad b_{kh}^k  = - p_{h,k} \, S \, (1-\tau_k) \, \frac{1}{r_{k+1} - 
r_{k}} \pb \nonumber
\label{bhk}
\end{eqnarray} 
Therefore, the possibly nonzero elements
$b_{hk}^i$ are of the form 
\begin{eqnarray}
b_{i+1,k}^{i} & = 
                  & p_{i+1,k} \, S \, (1-\tau_k) \, \frac{1}{r_{i+1} - r_{i}} \vb \nonumber \\
b_{i,k}^i & = 
            & - \, p_{k,i} \, S \, (1-\tau_i) \, \frac{1}{r_{i+1} - r_{i}} 
               - \, p_{i,k} \, S \, (1-\tau_k) \, \frac{1}{r_{i} - r_{i-1}} \vb \nonumber \\
b_{i-1,k}^i & = 
               & p_{k,i-1} \, S \, (1-\tau_{i-1}) \, \frac{1}{r_{i} - 
	       r_{i-1}} \vb
\label{b}
\end{eqnarray} 
where 
the expression for $b_{i+1,k}^{i}$ in $(\ref{b})$ holds true
for $i \le n-1$ and $k\le n-1$, 
in the expression for $b_{i,k}^i$,
the first addendum is effectively present only 
provided $i \le n-1$ and $k \ge 2$ and
the second addendum
only 
provided $i \ge 2$ and $k \le n-1$, 
and the expression for $b_{i-1,k}^i$ holds true for $i \ge 2$ and $k\ge 2$.

\smallskip

We express the indirect transition variation densities $T_{[hk]}^i(x)$ as
$$
T_{[hk]}^i(x) =  
U_{[hk]}^i(x) + V_{[hk]}^i(x)
\vb
$$
where 
\begin{equation}
U_{[hk]}^i(x) =  
\frac{p_{h,k} \, S \, \tau_k}{\sum_{j=1}^{n} x_{j}} {\bigg (}  \frac{x_{i-1}}{(r_i - r_{i-1})} -   \frac{x_{i}}{(r_{i+1} - r_{i})} {\bigg )}
\label{U_{[hk]}^i(x)}
\end{equation}
represents
the variation density corresponding to the advancement
from a class to the subsequent one, due to the benefit of taxation
and
\begin{equation}
V_{[hk]}^i(x)
=  p_{h,k} \, S \, \tau_k \, 
{\bigg (} 
\frac{\delta_{h,i+1}}{r_h - r_{i}} \, - \, \frac{\delta_{h,i}}{r_h - r_{i-1}}
{\bigg )} 
\, \frac{\sum_{j=1}^{n-1} x_{j}}{\sum_{j=1}^{n} x_{j}}
\vb
\label{V_{[hk]}^i(x)}
\end{equation}
with
$\delta_{h,k}$ denoting the 
{\slshape Kronecker delta},
accounts for 
the variation density corresponding to the retrocession
from a class to the preceding one, due to the payment
of some tax.
In the r.h.s. of 
$(\ref{U_{[hk]}^i(x)})$ and $(\ref{V_{[hk]}^i(x)})$,
$h >1$ 
and
the terms 
involving the index $i-1$ [respectively, $i+1$]
are effectively present only provided $i-1 \ge 1$ 
[respectively, $i+1 \le n$].

Notice that for technical reasons,
in the model under consideration,
the effective amount of money 
paid as tax
relative to an exchange of $S (1 - \tau_k)$
between two individuals
and then redistributed among classes
is given by $S \, \tau_k \,({\sum_{j=1}^{n-1} x_{j}})/{(\sum_{j=1}^{n} x_{j}})$
instead of $S \, \tau_k$.

As we shall see in the next subsection, a general theorem ensures
that 
the solutions $x =(x_1, \ldots , x_n)$ of interest
remain normalized during the evolution, i.e.\ $\sum_{j=1}^{n} x_{j}=1$. As a consequence, the expressions for $U_{[hk]}^i(x)$ and $V_{[hk]}^i(x)$ are simplified and we see 
from $(\ref{U_{[hk]}^i(x)})$ and $(\ref{V_{[hk]}^i(x)})$
that they both are linear functions of the $x_i$.

\subsection{Existence and uniqueness of the solution, conserved quantities, non-negativity and normalization property}

The well-posedness of the model was established in \cite{Ber}. Precisely, it was proven there that,
with the choice of parameters as in Subsection $2.2$, in correspondence to any initial non-negative and normalized condition
$x_{0}$, a unique non-negative and normalized solution 
$x(t)$ exists. In geometrical terms, 
taken an initial condition
$x_0 = (x_{01} , \ldots , x_{0n}) \in \Sigma_{n-1}$,
where 
\begin{equation}
\Sigma_{n-1} = 
{\Big \{} 
x = (x_1, ..., x_n) \in {\bf R}^n : x_{i} \ge 0 \ \hbox{for any} \
i \in \Gamma_n \ \hbox{and} \ \sum_{i=0}^n x_{i} = 1
{\Big \}}
\label{simplex}
\end{equation}
denotes the  \textquotedblleft $(n-1)$-simplex\textquotedblright,
a unique solution $x(t) = (x_1(t),\ldots,x_n(t))$ of $(\ref{evolution eq eta = 1})$ exists,
which is defined for all $t \in [0,+\infty)$ and satisfies $x(0) = x_0$. Moreover, 
\begin{equation}
x(t) \in \Sigma_{n-1} \quad \hbox{for all} \ t \ge 0 \pb 
\label{solution in the future}
\end{equation}
Only initial data $x_0$ on the $(n-1)$-simplex will be considered in this paper. Indeed, 
the requirement that $x_{0i} \ge 0$ for any $i \in \Gamma_n$ is totally natural, and the assumption
$\sum_{i=0}^n x_{i0} = 1$ expresses nothing but a normalization.
The validity of $(\ref{solution in the future})$ guarantees that the 
$x_i(t)$ for $i \in \Gamma_n$ are in fact the components of a distribution function and allows to further simplify 
the expressions of the $U_{[hk]}^i(x)$ and $V_{[hk]}^i(x)$ in 
$(\ref{U_{[hk]}^i(x)})$
and 
$(\ref{V_{[hk]}^i(x)})$
respectively.

We may from now on 
consider, instead of $(\ref{evolution eq eta = 1})$, the system of differential equations
\begin{equation}
{{d x_i} \over {d t}} =  
\sum_{h=1}^n \sum_{k=1}^n {\Big (} C_{hk}^i + T_{[hk]}^i(x) {\Big )}
x_h x_k     -    x_i  
\vb
\qquad i \in {\Gamma}_n \vb
\label{simplified evolution eq eta = 1}
\end{equation}
where the terms $T_{[hk]}^i(x)$ are linear in the variables $x_j$.
The equations in $(\ref{simplified evolution eq eta = 1})$ 
have a polynomial right hand side,
containing cubic terms as the highest degree ones.
 
Due to the fact that the value of $n$ and the parameters $r_k, \tau_k, p_{h,k}$ 
are still to be fixed,
the equations $(\ref{simplified evolution eq eta = 1})$ 
actually describe a family of models 
rather than a single model.

As proven in \cite{Ber}, the scalar function
$\mu(x)=\sum_{i=1}^n r_i x_i$,
which expresses the global wealth of the closed society under investigation, is conserved in the evolution, i.e.\ it
is a first integral for the system $(\ref{simplified evolution eq eta = 1})$. 
We also point out that, in view of the normalization of the population, the global wealth
coincides here with the mean wealth.
Taking advantage of the positive invariance
$(\ref{solution in the future})$ of the 
$(n-1)$-simplex $\Sigma_{n-1}$ and of the conservativity of $\mu(x)$, it is possible to reduce the dimension of the system 
to be studied.
Precisely, for any admissible value $\mu$ of 
the total wealth $\mu(x)$,
we are led to consider a system of $n-2$
nonlinear ordinary differential equations
with $n-2$ unknown functions. In fact,
we have a one-parameter family 
of systems of $n-1$ differential equations, $\mu \in [r_{1},r_{n}]$ being the parameter.

\section{Asymptotic stationary distributions}

Being interested in the long-time behavior of the solutions of the equations $(\ref{simplified evolution eq eta = 1})$,
we report in this section
on the outcomes of several computational simulations.
Of course, to carry out the simulations,
the value of $n$ has to be fixed, as well as the parameters of the model.

To start with, we 
take $S=1$ and
chose the elements of  the matrix $P$ to be all equal to $1/4$, apart from those lying 
on the main diagonal, on the first and the $n$-th row, on the first and the $n$-th column.
Those elements were taken to be given as
$$
\begin{array}{llll}
\ & \ & p_{h,h} = 1/2 \quad & \hbox{for} \ h \in \{2, ..., n-1\} \vb \nonumber \\
p_{1,k} = 0 \quad & \hbox{for} \ k \in \{1, ..., n\} \vb \quad & p_{n,k} = 1/2 \quad & \hbox{for} \ k \in \{1, ..., n\} \vb \nonumber \\
p_{h,n} = 0 \quad & \hbox{for} \ h \in \{1, ..., n-1\} \vb \quad & p_{h,1} = 1/2 \quad & \hbox{for} \ h \in \{2, ..., n\} \pb \nonumber 
\label{P}
\end{array} 
$$
Such a choice amounts to postulate that some money exchange takes place with probability $1/2$ on the occasion of every 
individual interaction and,
with exception for the interactions involving individuals of the first or of the last class, 
the probability that it's one or the other individual who pays is the same. 

To exploit the flexibility of the framework, we tried and work with
various 
choices of the 
values of $r_k$ and $\tau_k$.

\smallskip

Our findings are described in the next subsections.

\subsection{Uniqueness of the asymptotic stationary distribution for a fixed value $\mu$ of the total wealth}

Any time the value of $n$ and the parameters $r_k, \tau_k$
were fixed,
the simulations 
gave evidence of the following fact:
for any fixed value $\mu \in [r_1,r_n]$ of the global wealth,
an equilibrium - namely a stationary distribution -
exists, which coincides with the asymptotic trend of all solutions of $(\ref{simplified evolution eq eta = 1})$,
whose initial conditions $x_0 = (x_{01} , \ldots , x_{0n})$ belong to the 
\textquotedblleft $(n-1)$-simplex\textquotedblright 
(\ref{simplex}) and
satisfy 
$\sum_{i=1}^n r_i x_{0i} = \mu$.

\newpage

\begin{figure}[h]
  \begin{center}
  \includegraphics[width=4cm,height=2cm]  {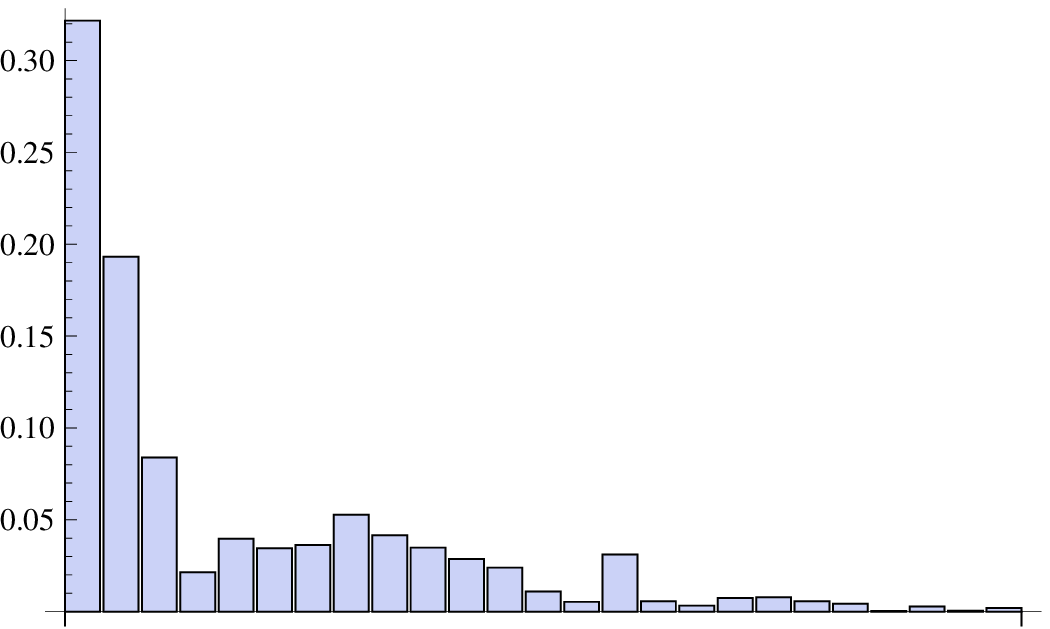}
  \hskip0.5cm
  \includegraphics[width=4cm,height=2cm]  {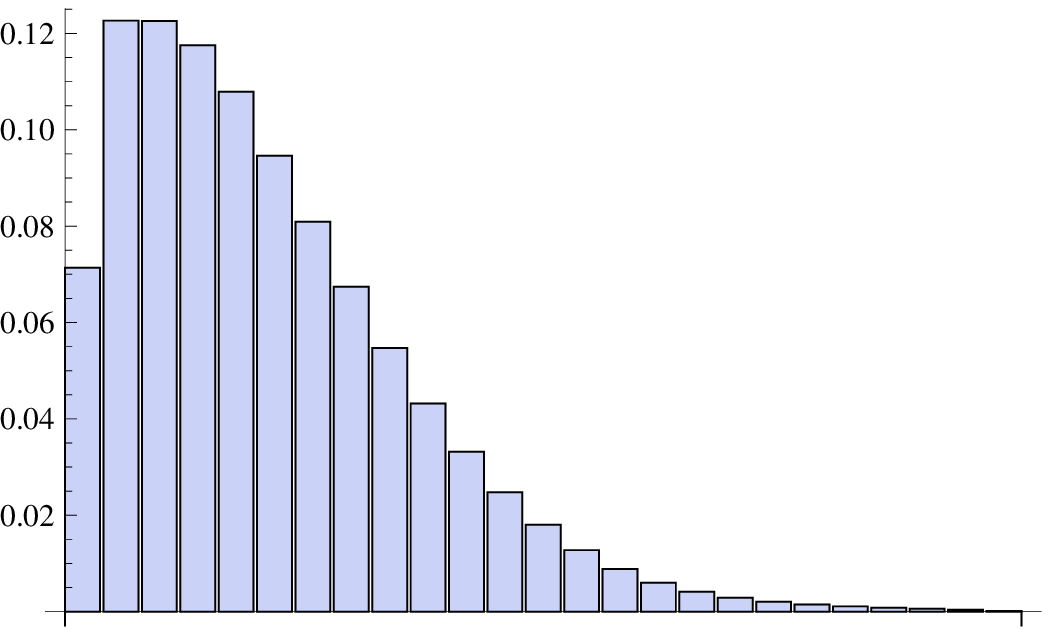}
  \end{center}
  \begin{center}
  \includegraphics[width=4cm,height=2cm]  {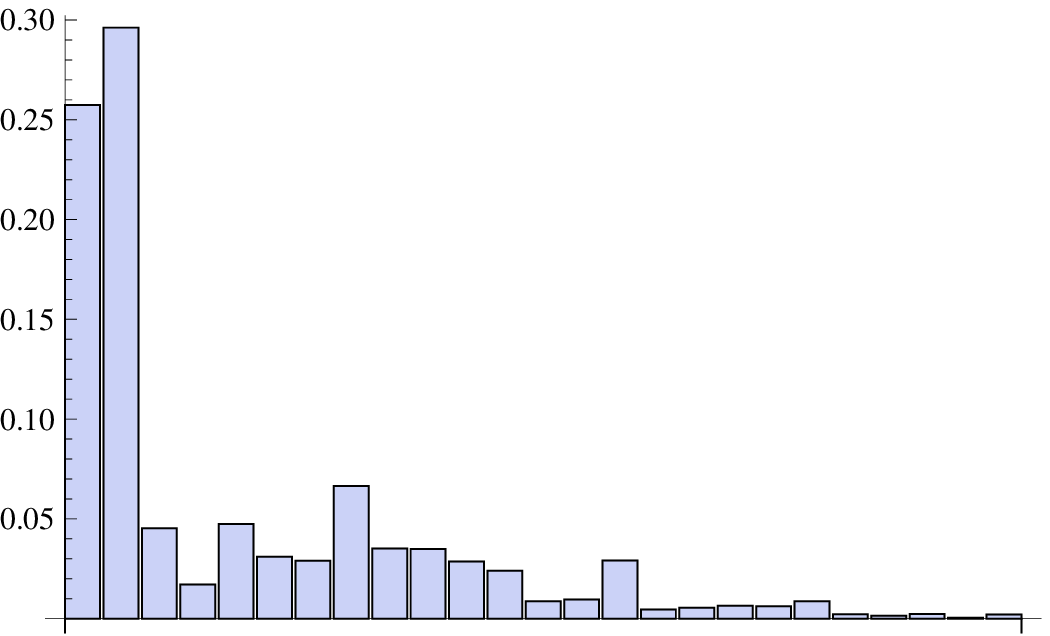}
  \hskip0.5cm
  \includegraphics[width=4cm,height=2cm]  {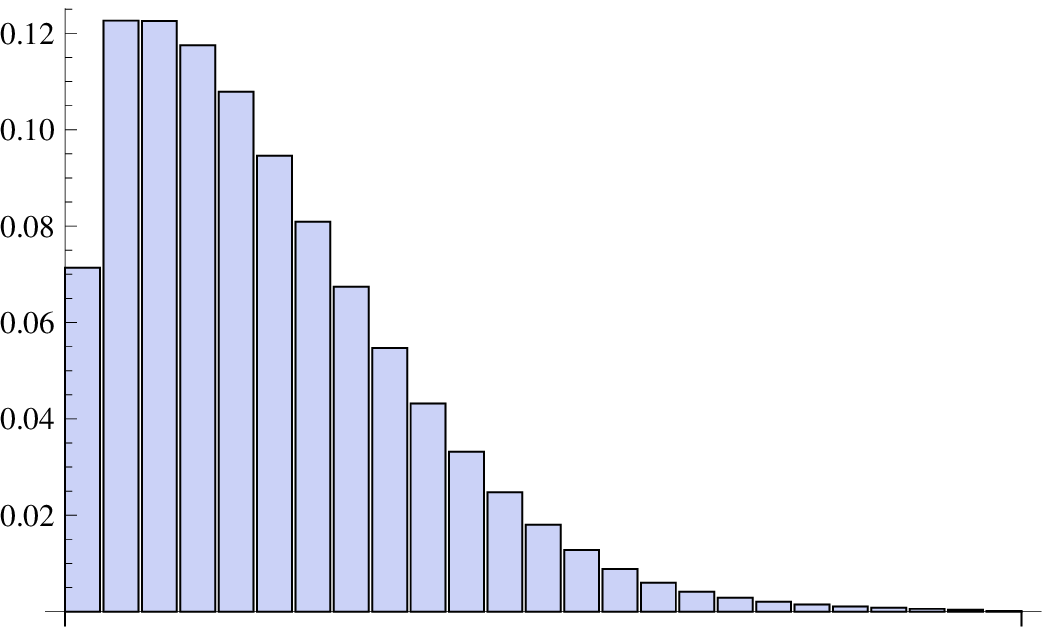}
  \end{center}
   \begin{center}
  \includegraphics[width=4cm,height=2cm]  {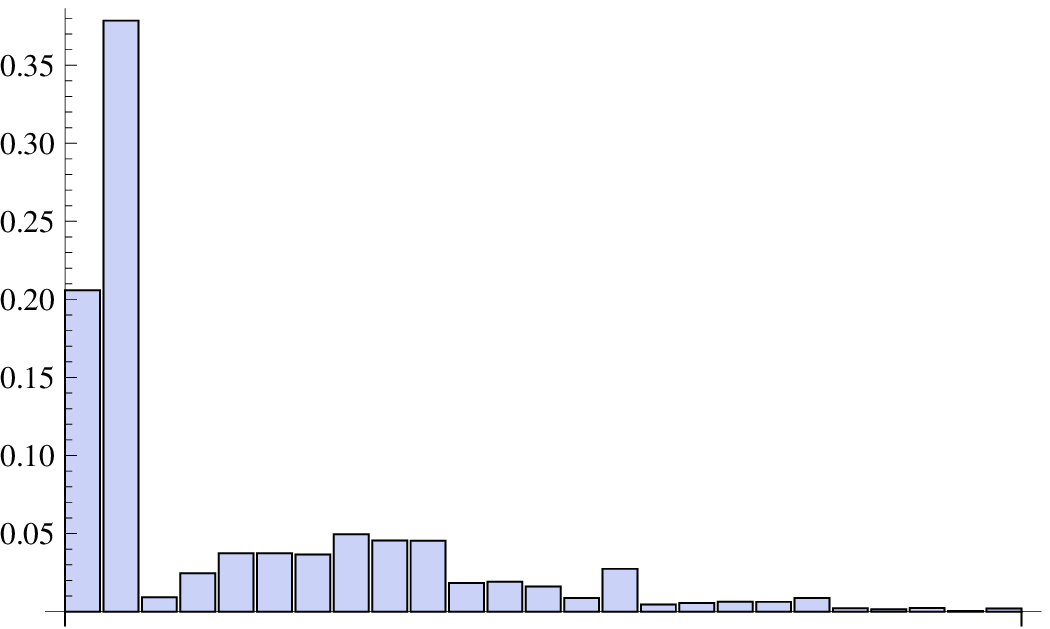}
  \hskip0.5cm
  \includegraphics[width=4cm,height=2cm]  {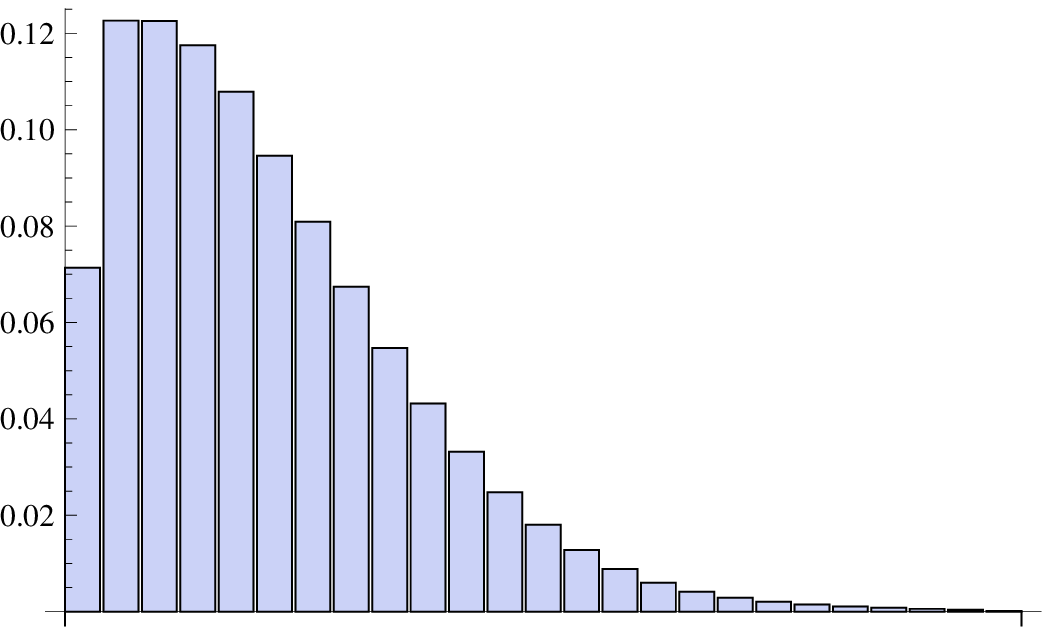}
  \end{center}
\caption{Initial (on the left) and long-time equilibrium (on the right) distributions for the model in Example $3.1.1$. 
Notice that
the histograms are scaled differently on different pictures.} 
\end{figure}

\vskip1.05truecm

\hrule
\begin{tabular}{llllllllll}
\                \ & \ & initial             \ & asymptotic   \ &\ &  initial             \ & asymptotic    \ &\ &  initial              \ & asymptotic   \nonumber \\
\                \ & \ & \                      \ & \                      \ &\ &  \                      \ & \                      \ &\ &  \                       \ & \                     \nonumber \\
$x_{1}$   \ & \ & $0.321732$ \ & $0.071378$ \ &\ &  $0.257386$ \ & $0.071371$ \ &\ &  $0.205908$ \ & $0.071368$ \nonumber \\
$x_{2}$   \ & \ & $0.193243$ \ & $0.122656$ \ &\ &  $0.296198$ \ & $0.122646$ \ &\ &  $0.378561$ \ & $0.122643$ \nonumber \\
$x_{3}$   \ & \ & $0.083916$ \ & $0.122567$ \ &\ &  $0.045308$ \ & $0.122560$ \ &\ &  $0.009230$ \ & $0.122558$ \nonumber \\
$x_{4}$   \ & \ & $0.021416$ \ & $0.117529$ \ &\ &  $0.017133$ \ & $0.117524$ \ &\ &  $0.024517$ \ & $0.117523$ \nonumber \\
$x_{5}$   \ & \ & $0.039693$ \ & $0.107903$ \ &\ &  $0.047480$ \ & $0.107901$ \ &\ &  $0.037439$ \ & $0.107901$ \nonumber \\
$x_{6}$   \ & \ & $0.034564$ \ & $0.094594$ \ &\ &  $0.031059$ \ & $0.094595$ \ &\ &  $0.037462$ \ & $0.094595$ \nonumber \\
$x_{7}$   \ & \ & $0.036266$ \ & $0.080894$ \ &\ &  $0.029012$ \ & $0.080896$ \ &\ &  $0.036663$ \ & $0.080898$ \nonumber \\
$x_{8}$   \ & \ & $0.052777$ \ & $0.067416$ \ &\ &  $0.066430$ \ & $0.067421$ \ &\ &  $0.049530$ \ & $0.067423$ \nonumber \\
$x_{9}$   \ & \ & $0.041546$ \ & $0.054706$ \ &\ &  $0.035146$ \ & $0.054711$ \ &\ &  $0.045536$ \ & $0.054714$ \nonumber \\
$x_{10}$ \ & \ & $0.034848$ \ & $0.043194$ \ &\ &  $0.034848$ \ & $0.043200$ \ &\ &  $0.045453$ \ & $0.043203$ \nonumber \\
$x_{11}$ \ & \ & $0.028657$ \ & $0.033173$ \ &\ &  $0.028657$ \ & $0.033179$ \ &\ &  $0.018357$ \ & $0.033180$ \nonumber \\
$x_{12}$ \ & \ & $0.023959$ \ & $0.024781$ \ &\ &  $0.023959$ \ & $0.024786$ \ &\ &  $0.019167$ \ & $0.024787$ \nonumber \\
$x_{13}$ \ & \ & $0.010926$ \ & $0.018017$ \ &\ &  $0.008741$ \ & $0.018022$ \ &\ &  $0.016124$ \ & $0.018023$ \nonumber \\
$x_{14}$ \ & \ & $0.005316$ \ & $0.012767$ \ &\ &  $0.009536$ \ & $0.012771$ \ &\ &  $0.008663$ \ & $0.012771$ \nonumber \\
$x_{15}$ \ & \ & $0.031142$ \ & $0.008837$ \ &\ &  $0.029107$ \ & $0.008839$ \ &\ &  $0.027389$ \ & $0.008839$ \nonumber \\
$x_{16}$ \ & \ & $0.005713$ \ & $0.005992$ \ &\ &  $0.004570$ \ & $0.005993$ \ &\ &  $0.004570$ \ & $0.005993$ \nonumber \\
$x_{17}$ \ & \ & $0.003274$ \ & $0.004126$ \ &\ &  $0.005494$ \ & $0.004127$ \ &\ &  $0.005494$ \ & $0.004126$ \nonumber \\
$x_{18}$ \ & \ & $0.007481$ \ & $0.002892$ \ &\ &  $0.006404$ \ & $0.002892$ \ &\ &  $0.006404$ \ & $0.002891$ \nonumber \\
$x_{19}$ \ & \ & $0.007814$ \ & $0.002064$ \ &\ &  $0.006251$ \ & $0.002062$ \ &\ &  $0.006251$ \ & $0.002062$ \nonumber \\
$x_{20}$ \ & \ & $0.005718$ \ & $0.001494$ \ &\ &  $0.008767$ \ & $0.001493$ \ &\ &  $0.008767$ \ & $0.001492$ \nonumber \\
$x_{21}$ \ & \ & $0.004265$ \ & $0.001091$ \ &\ &  $0.002223$ \ & $0.001090$ \ &\ &  $0.002223$ \ & $0.001089$ \nonumber \\
$x_{22}$ \ & \ & $0.000415$ \ & $0.000797$ \ &\ &  $0.001502$ \ & $0.000795$ \ &\ &  $0.001502$ \ & $0.000795$ \nonumber \\
$x_{23}$ \ & \ & $0.002820$ \ & $0.000576$ \ &\ &  $0.002289$ \ & $0.000574$ \ &\ &  $0.002289$ \ & $0.000574$ \nonumber \\
$x_{24}$ \ & \ & $0.000483$ \ & $0.000406$ \ &\ &  $0.000483$ \ & $0.000405$ \ &\ &  $0.000483$ \ & $0.000405$ \nonumber \\
$x_{25}$ \ & \ & $0.002017$ \ & $0.000147$ \ &\ &  $0.002017$ \ & $0.000147$ \ &\ &  $0.002017$ \ & $0.000147$ \nonumber  
\end{tabular}
\hrule 
\smallskip
Table 3.1.1
The initial and the asymptotic components of the three solutions in Figure $1$.

\newpage

\begin{figure}[h]
  \begin{center}
  \includegraphics[width=4cm,height=2cm]  {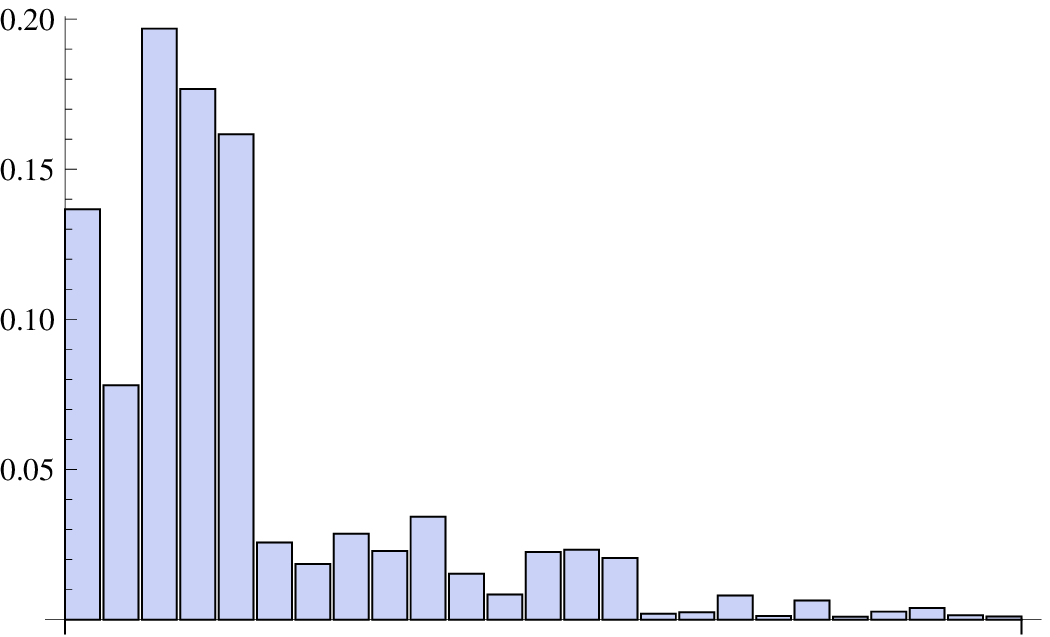}
  \hskip0.5cm
  \includegraphics[width=4cm,height=2cm]  {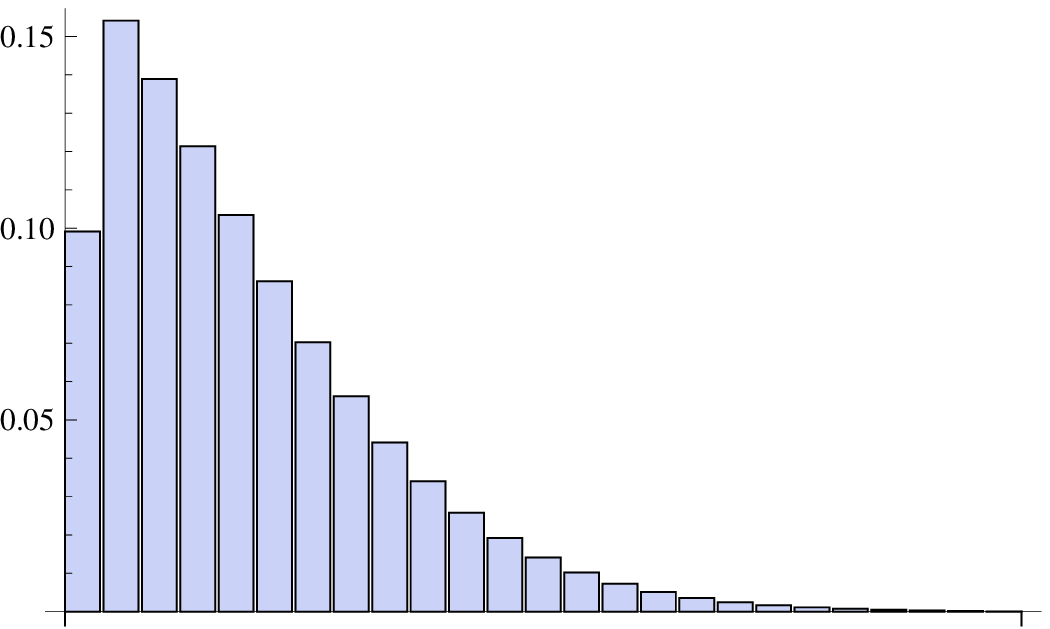}
  \end{center}
  \begin{center}
  \includegraphics[width=4cm,height=2cm]  {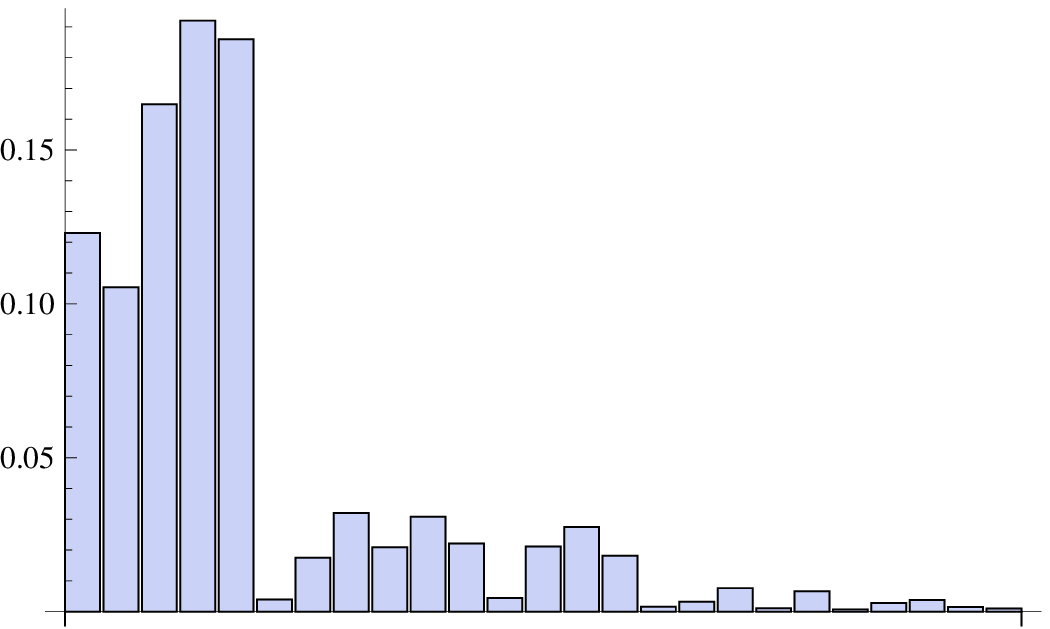}
  \hskip0.5cm
  \includegraphics[width=4cm,height=2cm]  {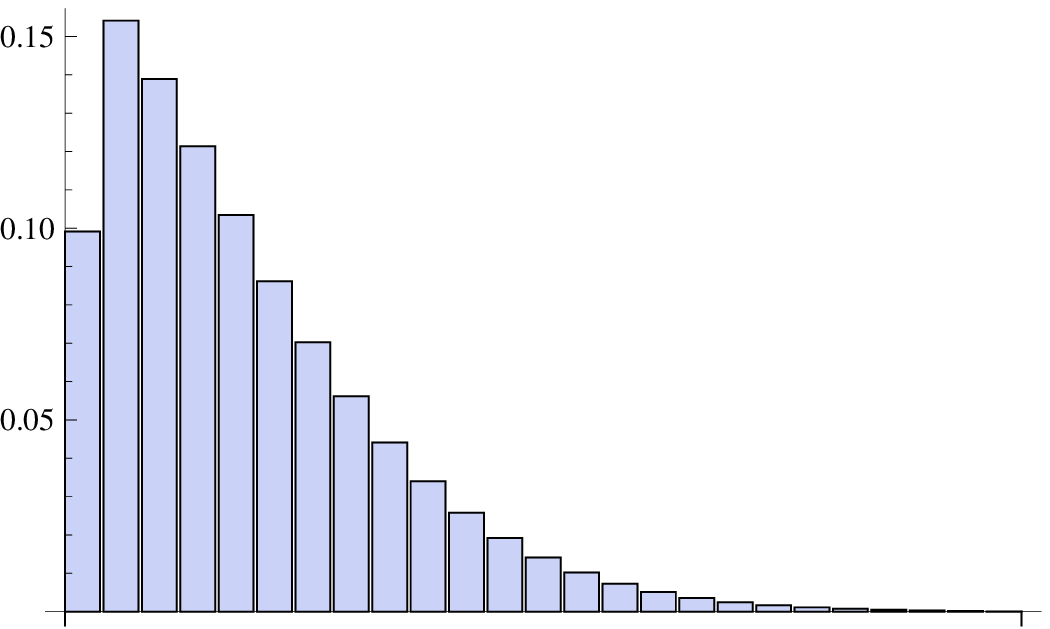}
  \end{center}
   \begin{center}
  \includegraphics[width=4cm,height=2cm]  {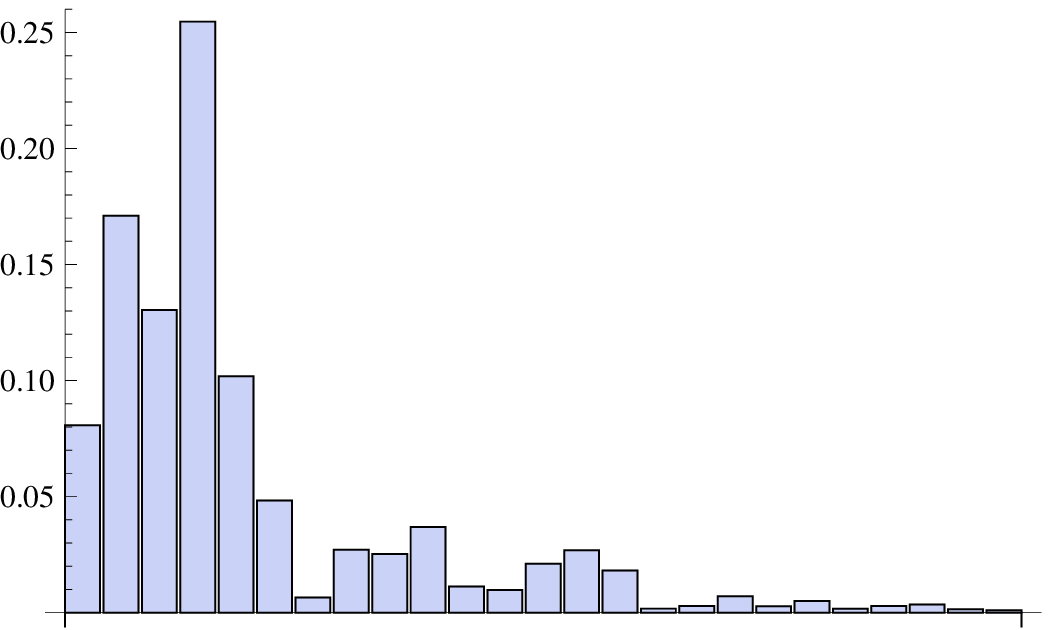}
  \hskip0.5cm
  \includegraphics[width=4cm,height=2cm]  {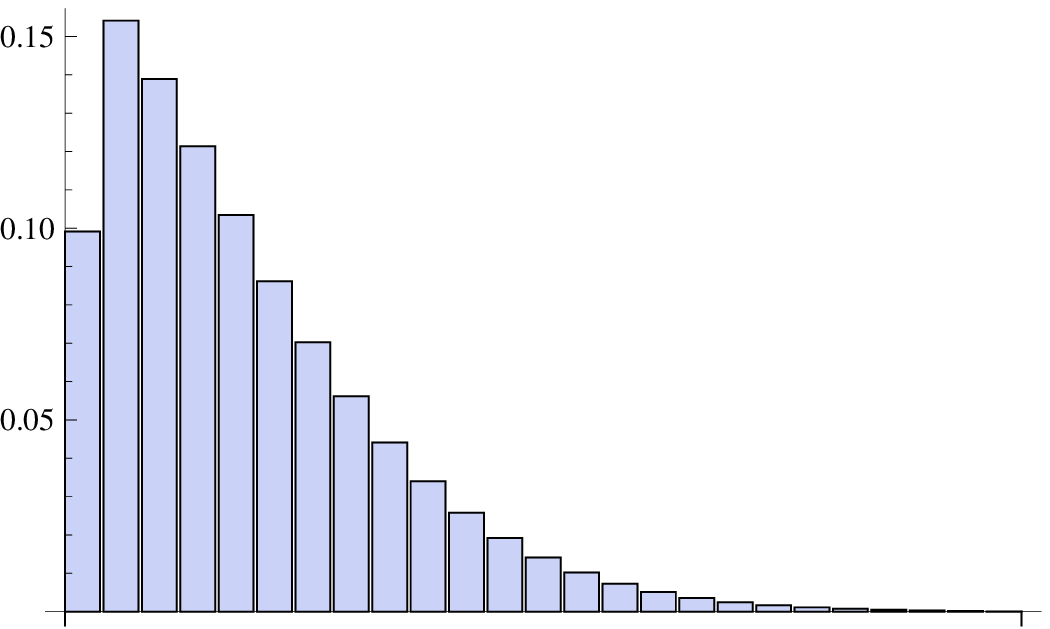}
  \end{center}
\caption{Initial (on the left) and long-time equilibrium (on the right) distributions for the model in Example $3.1.2$. 
Notice that
the histograms are scaled differently on different pictures.} 
\end{figure}

\vskip1.05truecm

\hrule
\begin{tabular}{llllllllll}
\                \ & \ & initial             \ & asymptotic    \ &\ &  initial             \ & asymptotic    \ &\ &  initial              \ & asymptotic   \nonumber \\
\                \ & \ & \                      \ & \                      \ &\ &  \                      \ & \                      \ &\ &  \                       \ & \                     \nonumber \\
$x_{1}$   \ & \ & $0.136714$ \ & $0.099140$ \ & \ & $0.123043$ \ & $0.099140$ \ & \ & $0.080728$ \ & $0.099140$ \nonumber \\
$x_{2}$   \ & \ & $0.078092$ \ & $0.154143$ \ & \ & $0.105435$ \ & $0.154143$ \ & \ & $0.171057$ \ & $0.154143$ \nonumber \\
$x_{3}$   \ & \ & $0.196837$ \ & $0.138916$ \ & \ & $0.164849$ \ & $0.138916$ \ & \ & $0.130391$ \ & $0.138916$ \nonumber \\
$x_{4}$   \ & \ & $0.176736$ \ & $0.121397$ \ & \ & $0.192032$ \ & $0.121397$ \ & \ & $0.254674$ \ & $0.121397$ \nonumber \\
$x_{5}$   \ & \ & $0.161621$ \ & $0.103429$ \ & \ & $0.185978$ \ & $0.103429$ \ & \ & $0.101834$ \ & $0.103429$ \nonumber \\
$x_{6}$   \ & \ & $0.025700$ \ & $0.086143$ \ & \ & $0.003927$ \ & $0.086143$ \ & \ & $0.048330$ \ & $0.086143$ \nonumber \\
$x_{7}$   \ & \ & $0.018566$ \ & $0.070253$ \ & \ & $0.017495$ \ & $0.070253$ \ & \ & $0.006526$ \ & $0.070253$ \nonumber \\
$x_{8}$   \ & \ & $0.028615$ \ & $0.056173$ \ & \ & $0.032066$ \ & $0.056173$ \ & \ & $0.027059$ \ & $0.056173$ \nonumber \\
$x_{9}$   \ & \ & $0.022840$ \ & $0.044083$ \ & \ & $0.020896$ \ & $0.044083$ \ & \ & $0.025315$ \ & $0.044083$ \nonumber \\
$x_{10}$ \ & \ & $0.034279$ \ & $0.033987$ \ & \ & $0.030850$ \ & $0.033987$ \ & \ & $0.036897$ \ & $0.033987$ \nonumber \\
$x_{11}$ \ & \ & $0.015286$ \ & $0.025764$ \ & \ & $0.022142$ \ & $0.025764$ \ & \ & $0.011226$ \ & $0.025764$ \nonumber \\
$x_{12}$ \ & \ & $0.008382$ \ & $0.019219$ \ & \ & $0.004459$ \ & $0.019219$ \ & \ & $0.009789$ \ & $0.019218$ \nonumber \\
$x_{13}$ \ & \ & $0.022512$ \ & $0.014118$ \ & \ & $0.021153$ \ & $0.014118$ \ & \ & $0.021096$ \ & $0.014118$ \nonumber \\
$x_{14}$ \ & \ & $0.023277$ \ & $0.010220$ \ & \ & $0.027482$ \ & $0.010220$ \ & \ & $0.026877$ \ & $0.010220$ \nonumber \\
$x_{15}$ \ & \ & $0.020542$ \ & $0.007294$ \ & \ & $0.018192$ \ & $0.007294$ \ & \ & $0.018198$ \ & $0.007294$ \nonumber \\
$x_{16}$ \ & \ & $0.001924$ \ & $0.005136$ \ & \ & $0.001558$ \ & $0.005136$ \ & \ & $0.001731$ \ & $0.005136$ \nonumber \\
$x_{17}$ \ & \ & $0.002448$ \ & $0.003570$ \ & \ & $0.003179$ \ & $0.003570$ \ & \ & $0.002833$ \ & $0.003570$ \nonumber \\
$x_{18}$ \ & \ & $0.008008$ \ & $0.002450$ \ & \ & $0.007642$ \ & $0.002450$ \ & \ & $0.007034$ \ & $0.002450$ \nonumber \\
$x_{19}$ \ & \ & $0.001222$ \ & $0.001660$ \ & \ & $0.001099$ \ & $0.001660$ \ & \ & $0.002785$ \ & $0.001660$ \nonumber \\
$x_{20}$ \ & \ & $0.006399$ \ & $0.001112$ \ & \ & $0.006643$ \ & $0.001112$ \ & \ & $0.005056$ \ & $0.001112$ \nonumber \\
$x_{21}$ \ & \ & $0.000938$ \ & $0.000736$ \ & \ & $0.000734$ \ & $0.000736$ \ & \ & $0.001670$ \ & $0.000736$ \nonumber \\
$x_{22}$ \ & \ & $0.002667$ \ & $0.000481$ \ & \ & $0.002830$ \ & $0.000481$ \ & \ & $0.002889$ \ & $0.000481$ \nonumber \\
$x_{23}$ \ & \ & $0.003895$ \ & $0.000311$ \ & \ & $0.003813$ \ & $0.000311$ \ & \ & $0.003503$ \ & $0.000311$ \nonumber \\
$x_{24}$ \ & \ & $0.001475$ \ & $0.000199$ \ & \ & $0.001475$ \ & $0.000199$ \ & \ & $0.001475$ \ & $0.000199$ \nonumber \\
$x_{25}$ \ & \ & $0.001025$ \ & $0.000069$ \ & \ & $0.001025$ \ & $0.000069$ \ & \ & $0.001025$ \ & $0.000069$ \nonumber
\end{tabular}
\hrule 
\smallskip
Table 3.1.2
The initial and the asymptotic components of the three solutions in Figure $2$.

\newpage

Indeed, selecting, for a fixed $\mu \in [r_1,r_n]$, 
various initial conditions 
$x_0 \in \Sigma_{n-1}$
for which $\sum_{i=1}^n r_i x_{0i} = \mu$ hold true,
we obtained results of the kind illustrated in the Figures $1$ and $2$.

\medskip

{\it{Example $3.1.1$.}} Take $n=25$, the average incomes quadratically growing: $r_j = 10 \cdot j^2$ for $j = 1, ..., 25$ 
and let the vector with tax rates components $\tau_j$ with $j = 1, ..., 25$ be given by
({0, 0.05, 0.1, 0.15, 0.2, 0.225, 0.25, 0.275, 0.3, 0.325, 0.35, 0.375, 0.4, 0.425, 0.45, 0.45, 0.45, 0.45, 0.45, 0.45, 0.45, 0.45, 0.45, 0.45, 0.45}).
E.g., the three solutions evolving from the initial distributions on the left in Figure $1$ tend in the future to the corresponding distributions on the right, which essentially coincide.
The numerical values of the initial and the asymptotic components of the three solutions are given in the Table $3.1.1$.
The constant value of the global wealth along the three solutions is $\mu =483$. 

\medskip

{\it{Example $3.1.2$.}} Take $n=25$, the average incomes linearly growing: $r_j = 10 \cdot j$ for $j = 1, ..., 25$ 
and take the tax rate $\tau_j = \tau_{min} + (\tau_{max} -  \tau_{min}) \sqrt{(j-1)/24}$, where $\tau_{min}=0$ and $\tau_{max}=40/100$, for $j = 1, ..., 25$.
E.g., the three solutions evolving from the initial distributions on the left in Figure $1$ tend in the future to the corresponding distributions on the right, which essentially coincide.
The numerical values of the initial and the asymptotic components of the three solutions are given in the Table $3.1.2$.
The constant value of the global wealth along the three solutions is $\mu =52.52$. 

\subsection{Dependence of the asymptotic stationary distribution on the total wealth $\mu$}

For fixed $n$ and fixed $r_k$ and $\tau_k$
we considered a list of different initial distributions corresponding to different values of the global wealth.
We focused on long-time
numerical solutions evolving from these distributions.
Then, we fixed other values of $n$, of $r_k$ and $\tau_k$
and
repeated the test.

We drew the
conclusion that for a fixed model, the outline of the asymptotic stationary distribution depends on the conserved quantity $\mu$,
which is related to the initial condition. In other words, we may say that there is a one-parameter family of asymptotic stationary distributions.

In this connection, we also want to emphasize 
a point, 
which will become clearer in the Subsection $3.4$.
Other elements which, together with the value of global wealth $\mu$,
have a decisive
effect
on the shape of the asymptotic distributions 
are
the fact that in the framework described in this paper (only) a finite number of income classes are scheduled
and the fact that, by (\ref{solution in the future}), the number of individuals remains constant too.

\subsection{Dependence of the asymptotic stationary distribution on the difference between the maximum and the minimum tax rate}

Next we compare one with another different models.

In particular,  for a fixed choice of
$n$ and $r_k$
and for fixed total wealth $\mu$, we put to test
different tax rates $\tau_k$. When doing that, it can be observed that
the outline of the asymptotic stationary distribution depends on
the difference between the maximum
and the minimum tax rate, i.e. the rates
respectively applied to the highest and to the lowest  
income classes. Specifically,
to
an increase of the difference between the maximum and the minimum tax rates $\tau_{max}$ and $\tau_{min}$, 
a growth of the middle classes at the asymptotic equilibrium corresponds,
to the detriment of the poorest and the richest classes (see also \cite{Ber}).
For an illustrative purpose, we just report here the findings relative to one specific case.

\medskip

{\it{Example $3.3.1$.}} Take $n=25$, 
the average incomes linearly growing: $r_j = 10 \cdot j$ for $j = 1, ..., 25$ 
and take the tax rate $\tau_j = \tau_{min} + (\tau_{max} -  \tau_{min}) \sqrt{(j-1)/24}$. 
The Table $3.3.1$ reports the components of the asymptotic distributions 
corresponding to a same initial distribution
for the three models in which the 
minimum and the maximum tax rate respectively are: 
$$
\begin{array}{lll}
\hbox{case i)} :
& \tau_{min}=20\% \vb
& \tau_{max}=40\% \vb \nonumber \\
\hbox{case ii)} :
& \tau_{min}=10\% \vb
& \tau_{max}=50\% \vb \nonumber \\
\hbox{case iii)} :
& \tau_{min}=0\% \vb
& \tau_{max}=60\% \pb \nonumber 
\label{different tax system}
\end{array} 
$$
In the three cases the value of the total wealth is $\mu =50.43$.

Looking at the Table $3.3.1$, it is immediate noticing that the individual density in the first two classes, as well as in the last fourteen ones
is smaller when the difference $\tau_{max} -  \tau_{min}$ is larger. 
The reverse property does not hold true component-by-component
for the middle classes. However, for them
a collective property occurs: indeed,
the total density of the  of classes from the third to the $11$-th one is larger for larger $\tau_{max} -  \tau_{min}$.

\vskip1.05truecm 


\hrule
\begin{tabular}{lccc}
\  \ & case i) : 20\%-40\% \ & case ii) : 10\%-50\% \ & case iii) : 0\%-60\%  \nonumber \\
\quad\quad\quad\quad\quad\quad\quad & \ \ & \ \ & \ \ \nonumber \\
$x_{1}$ \ & $0.13261$ \ & $0.11637,$ \ & $0.09613$ \nonumber \\
$x_{2}$ \ & $0.16666$ \ & $0.16058$ \ & $0.15558$ \nonumber \\
$x_{3}$ \ & $0.13938$ \ & $0.13983$ \ & $0.14399$ \nonumber \\
$x_{4}$ \ & $0.11544$ \ & $0.11975$ \ & $0.12761$ \nonumber \\
$x_{5}$ \ & $0.09470$ \ & $0.10079$ \ & $0.10913$ \nonumber \\
$x_{6}$ \ & $0.07694$ \ & $0.08333$ \ & $0.09041$ \nonumber \\
$x_{7}$ \ & $0.06191$ \ & $0.06765$ \ & $0.07271$ \nonumber \\
$x_{8}$ \ & $0.04935$ \ & $0.05391$ \ & $0.05686$ \nonumber \\
$x_{9}$ \ & $0.03896$ \ & $0.04216$ \ & $0.04329$ \nonumber \\
$x_{10}$ \ & $0.03048$ \ & $0.03234$ \ & $0.03212$ \nonumber \\
$x_{11}$ \ & $0.02362$ \ & $0.02434$ \ & $0.02325$ \nonumber \\
$x_{12}$ \ & $0.01814$ \ & $0.01797$ \ & $0.01643$ \nonumber \\
$x_{13}$ \ & $0.01378$ \ & $0.01301$ \ & $0.01134$ \nonumber \\
$x_{14}$ \ & $0.01041$ \ & $0.00924$ \ & $0.00766$ \nonumber \\
$x_{15}$ \ & $0.00778$ \ & $0.00643$ \ & $0.00506$ \nonumber \\
$x_{16}$ \ & $0.00576$ \ & $0.00439$ \ & $0.00327$ \nonumber \\
$x_{17}$ \ & $0.00423$ \ & $0.00294$ \ & $0.00207$ \nonumber \\
$x_{18}$ \ & $0.00308$ \ & $0.00192$ \ & $0.00128$ \nonumber \\
$x_{19}$ \ & $0.00222$ \ & $0.00124$ \ & $0.00078$ \nonumber \\
$x_{20}$ \ & $0.00158$ \ & $0.00078$ \ & $0.00046$ \nonumber \\
$x_{21}$ \ & $0.00112$ \ & $0.00048$ \ & $0.00027$ \nonumber \\
$x_{22}$ \ & $0.00079$ \ & $0.00029$ \ & $0.00015$ \nonumber \\
$x_{23}$ \ & $0.00055$ \ & $0.00017$ \ & $0.00009$ \nonumber \\
$x_{24}$ \ & $0.00038$ \ & $0.00010$ \ & $0.00005$ \nonumber \\
$x_{25}$ \ & $0.00015$ \ & $0.00003$ \ & $0.00001$ \nonumber
\end{tabular}
\hrule 
\smallskip
Table 3.3.1
The components of the three asymptotic equilibria corresponding to three different taxation systems as described in the Example $3.3.1$.

\medskip

\subsection{Power-law distribution tails and dependence of the Pareto index on the total wealth $\mu$}

Aiming at a deeper analysis of the properties of the asymptotic stationary distributions,
we 
restricted attention on initial conditions, i.e. on initial distributions of the population,
having the majority of individuals concentrated in lower income classes.
To be concrete, we prescribed e.g. that 
a high percentage of individuals 
(in the example below, for example, $90 \%$)
belongs to the first five classes, 
another smaller percentage belongs to the second five classes, and so on. 
Such a restriction is motivated and justified by the fact that similar situations typically occur in real world societies.
Using words of \cite{Hee Mau},
we observe
that
non stationary initial distribution of individuals could be found
\textquotedblleft after a change in policy,
e.g. after a change in the income tax schedule, or during a demographic transition, 
as many modern industrialized countries experience it right now\textquotedblright.
We also point out in this connection that,
since in our models
the number of individuals does not change in time and
only a finite number $n$ of 
income classes is considered,
if a tail is expected in the asymptotic distribution,
the global wealth cannot be too high.
And this is obviously related to the mentioned realistic restriction.

The histograms of the asymptotic distributions seem to exhibit
a power-law behavior of the distribution tails. 
Before showing that this behavior actually occurs
and to see how to calculate the Pareto index \cite{Par},
we make here a short digression on the 
relation between the discrete distribution of components $x_i$ and the continuous income distribution $f(w)$.
We skip the variable $t$ since we have in mind here the asymptotic stationary distribution.

\medskip

Consider an income interval $(w_0,w_0+\Delta w)$, so small that the stationary distribution function $f(w)$ varies slowly in this interval, 
and yet such as to contain several discrete income classes $r(i)$, $r(i+1)$, ..., $r(i+\Delta i)$.
The income classes are chosen in such a way that $w_0=r(i)$ and $w_0+\Delta w=r(i+\Delta i)$. 
(Notice that $i$ is an integer variable and we usually denote the $i$-th income class with $r_i$; however, here we consider 
$r$ as a function of $i$, for reasons which will immediately be clear).

The population of the classes $i, ..., i+\Delta i$ will also vary slowly, and 
we can write the total population in this interval as
$$
f(w_0) \, \Delta w = x_i \, \Delta i \pb
$$
From this we see that the distribution function at $w_0$ is
$$
f(w_0) = x_i \, \frac{\Delta i}{\Delta w} \pb
$$
But $\Delta w = r(i+\Delta i)-r(i) \simeq r'(i) \, \Delta i$, so we obtain
$$
f(w_0) \simeq  \frac{x_i}{r'(i)} \pb
$$
Generalizing to any $w$ such that $w=r(i)$, we have
$$
f(w) \simeq  \frac{x_i}{r'(i)_{|_{i=r^{-1}(w)}}} \pb
$$
For instance, if the classes are chosen so that the income increases quadratically, namely
$$
r(i)=c i^2 
$$
for some constant $c>0$,
then the distribution function with components $x_i$ corresponding to the equilibrium solution of our evolution equations is, at each point $w=r(i)$,
$$
f(w) \simeq  \frac{x_i}{2\sqrt{c w}} \pb
$$

The power-law behavior $f(w) = {C} {w^{-\alpha}}$ 
is equivalently expressed by the equation of a straight line: $\ln f(w) = \ln C - \alpha \ln w$.
Therefore, 
to check whether the distribution tail
exhibits it and to compute the Pareto index, we must have a linear fit of the log-log plot in the variables $w$ and $f$.
What is known as Pareto index is the number $\beta = \alpha - 1$.

According to a quantity of empirical data relative to several countries with capitalistic economies, 
Pareto assessed the index $\beta$ to be approximately equal to $1.5$. More generally, $\beta$ was observed to take values between $1$ and $2$.

\bigskip

As for the models described in this paper, the power-law property has been checked in several cases, namely for several different choices of the parameters.
In particular, the following example illustrates 
the power-law behavior of the asymptotic distribution tail for a specific model. 

\medskip

\begin{figure}[h]
  \begin{center}
  \includegraphics[width=4cm,height=2cm]  {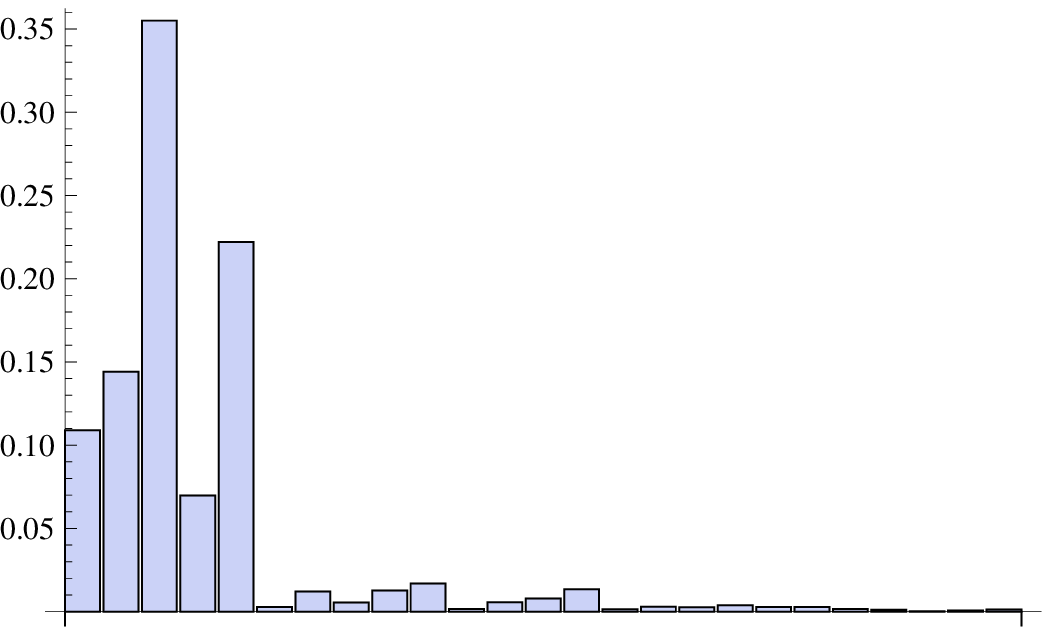}
  \hskip0.2cm
  \includegraphics[width=4cm,height=2cm]  {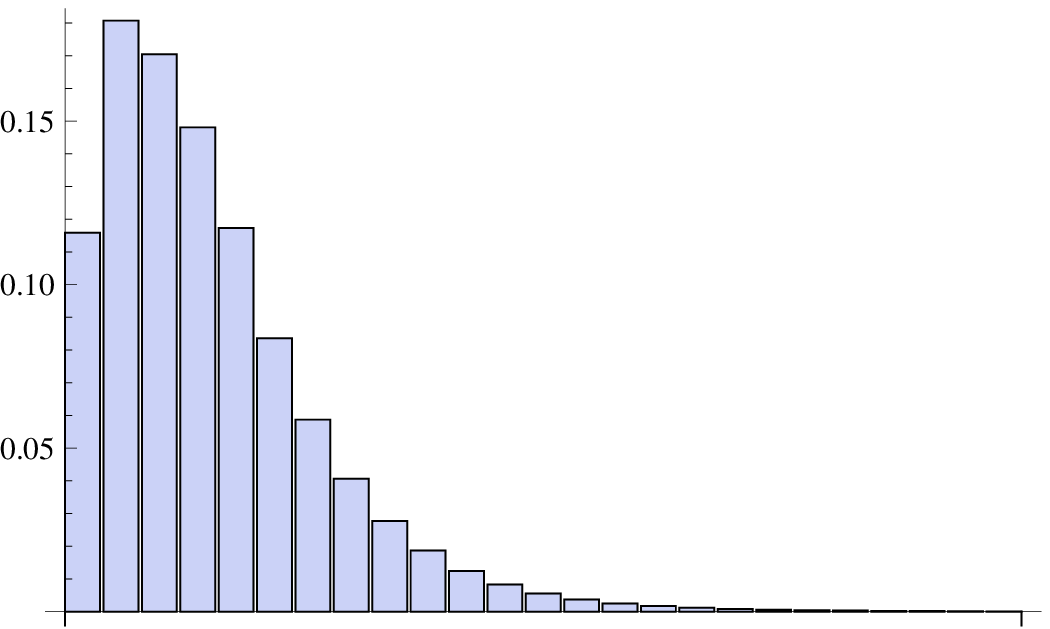}
   \hskip0.2cm
   \includegraphics[width=4cm,height=2cm]  {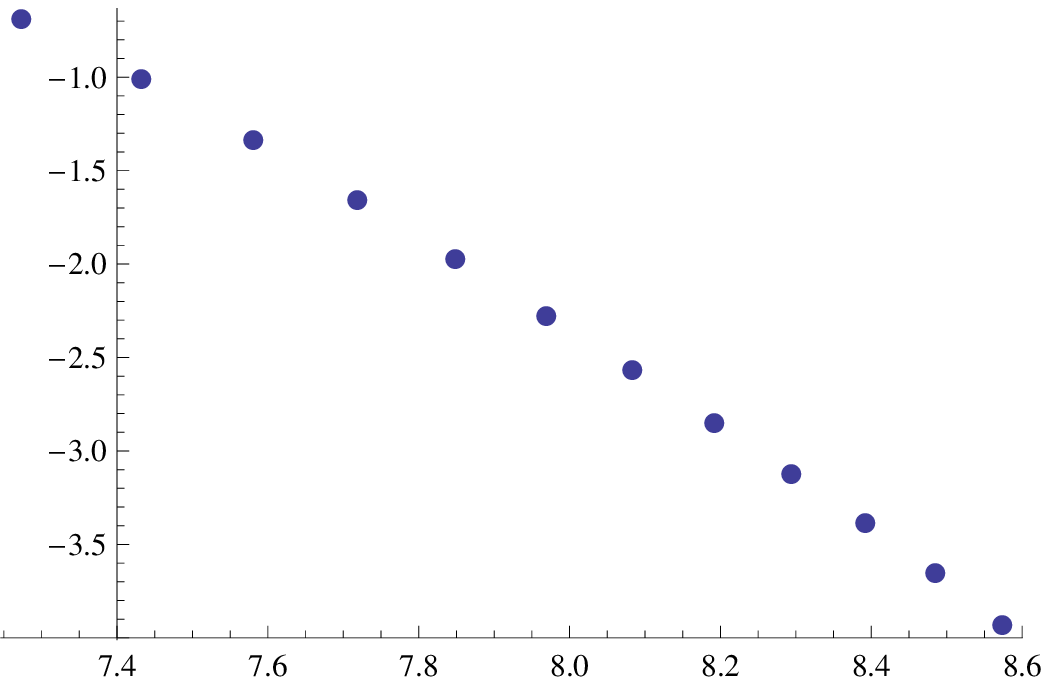}  
  \end{center}
\caption{The left and the central panel respectively illustrate an initial and its corresponding long-time equilibrium distribution for the model in Example $3.4.1$. In the right panel the log-log plot is reported. Similarly as in previous figures, the histograms are scaled differently on the left and the central panel.}
\end{figure}

\medskip

{\it{Example $3.4.1$.}} 
Take $n=25$, the average incomes quadratically growing: $r_j = 10 \cdot j^2$ for $j = 1, ..., 25$ 
and let the vector with tax rates components $\tau_j$ with $j = 1, ..., 25$ be given by
({0, 0.1, 0.2, 0.3, 0.4, 0.41, 0.42, 0.43, 0.44, 0.45, 0.45, 0.45, 0.45, 0.45, 0.45, 0.45, 0.45, 0.45, 0.45, 0.45, 0.45, 0.45, 0.45, 0.45, 0.45}).
The histograms on the left panel in Figure $3$ correspond to an initial distribution, whose asymptotic distribution is in the central panel. In the panel on the right
the log-log plot is illustrated, consisting of the sequence of points $(\ln (r_i), \ln (x_i/2 \sqrt{(10 r_i)}))$. 
The value of the total wealth is in this case $\mu = 271$ and the Pareto index is $1.7$.

\medskip

A further property, we want to emphasize is the seemingly monotone dependence of the Pareto index, and equivalently of the exponent $\alpha$, on the value 
$\mu$ of the total wealth. We tested this property on several different models. Here
we illustrate it with reference to the model described in the Example $3.4.1$:
we next list
a discrete number of values $\mu$ of the total wealth
together with the 
values $\alpha$ of the corresponding power-law exponents. 
The values in the list show
that $\alpha$ (and hence, also the Pareto index) decreases as $\mu$ increases.
\begin{center}
\begin{tabular}{|c|c|}
\hline
$\mu$ \ & $\alpha$ \ \nonumber \\
\hline
\ \ & \ \ \nonumber \\
$231.5$ \ & $2.81293$ \ \nonumber \\
$258.0$ \ & $2.76824$ \ \nonumber \\
$284.5$ \ & $2.70829$ \ \nonumber \\
$309.0$ \ & $2.63832$ \ \nonumber \\
$337.5$ \ & $2.56059$ \ \nonumber \\
$380.5$ \ & $2.39970$ \ \nonumber \\
$400.0$ \ & $2.32653$ \ \nonumber \\
$428.5$ \ & $2.23525$ \ \nonumber \\
$463.0$ \ & $2.11647$ \ \nonumber \\
$497.0$ \ & $1.96068$ \ \nonumber \\
\hline
\end{tabular}
\end{center}

\subsection{The Lorenz curve and the Gini coefficient}

The Gini coefficient is commonly (and not only) used as a measure of inequality of income or wealth. 
It can range from the value $0$, expressing complete equality, to the value $1$, expressing maximal inequality. 
It is obtained based on the Lorenz curve, which plots the cumulative
percentage of the total income of a population (on the $y$ axis) earned by the bottom percentage of individuals (on the $x$ axis).
In comparison with it, the line at $45$ degrees represents a perfect equality of incomes.
The Gini coefficient is defined as the ratio of
the area between the Lorenz curve and the line of perfect equality
and
the total area under the line of perfect equality.

\medskip

{\it{Example $3.5.1$.}} 
Consider the model described in the Example $3.4.1$. In Figure $4$ 
the line of perfect equality and the Lorenz curve relative to the asymptotic income distribution with total wealth $\mu=202$
are shown.

\medskip

\begin{figure}[h]
  \begin{center}
  \includegraphics[width=5cm,height=5cm]  {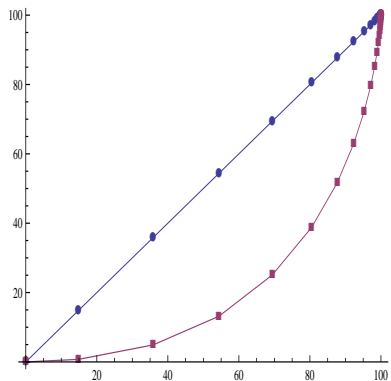}
  \end{center}
\caption{The line of perfect equality and the Lorenz curve relative to the asymptotic income distribution with total wealth $\mu=202$.}
\end{figure}

\medskip

Here are the values of 
the Gini coefficient $i_G$,
corresponding to the asymptotic distributions with a discrete choice of values of the total wealth $\mu$.
To estimate them, we calculated the area under the Lorenz curve as a sum of areas of trapezia.
Our purpose here is simply to show that the outputs of the family of models discussed exhibit possible agreement with real world data.
Gini indices as those reported below correspond for example to empirical data relative to the income in Brazil \cite{ChaFig Mou Rib}.
\begin{center}
\begin{tabular}{|c|c|}
\hline
$\mu$ \ & $i_G$ \ \nonumber \\
\hline
\ \ & \ \ \nonumber \\
$202$ \ & $0.588604$ \ \nonumber \\
$226$ \ & $0.592764$ \ \nonumber \\
$244$ \ & $0.596331$ \ \nonumber \\
$259$ \ & $0.598582$ \ \nonumber \\
$279$ \ & $0.601564$ \ \nonumber \\
\hline
\end{tabular}
\end{center}

\section{Conclusions}

A general framework is here discussed, within which the taxation and redistribution process in a closed society 
can be described and the evolution of the income distribution analyzed. The framework is expressed by 
a system of nonlinear ordinary differential equations, in which a number of parameters appear.
Fixing the parameter values amounts to formulating a specific model.
Relaxation to a stationary distribution exhibiting a Pareto type tail has been 
found in several cases. 
Although rough and naive, the proposed models allow a possible explanation of the shape of income distributions
commonly observed in 
real world 
economies.
In a further perspective, the framework is suitable for the construction of explorative models.
For example, it could be employed towards understanding which taxation system would be more desirable.

Future investigations should try to take one step further 
incorporating 
in the models, beside money exchanges, also assets and material possessions.
Also, it would be of interest somehow keeping into account tax evasion phenomena and checking their effects on the 
income and wealth distributions.







\end{document}